\documentclass[structabstract]{aa}  
\usepackage{graphicx}
\usepackage{url}
\usepackage{txfonts}
\begin{document}

   \title{The spectral index image of the radio halo in the cluster Abell 520 hosting a famous bow shock}

   \subtitle{}
\author{  V. Vacca\inst{1,2}
          \and
          L. Feretti\inst{2}
          \and
          G. Giovannini\inst{1,2}
          \and
          F. Govoni\inst{3}
          \and
          M. Murgia\inst{3}
          \and
          R. A. Perley\inst{4}
          \and
          T. E. Clarke\inst{5}
                   }
\institute{
              Dipartimento di Fisica e Astronomia, Universit\`a di Bologna, Via Ranzani 1, I--40127 Bologna, Italy
              \and   
              INAF - Istituto di Radioastronomia, Via Gobetti 101, I--40129 Bologna, Italy
              \and
              INAF - Osservatorio Astronomico di Cagliari, Via della Scienza 5, 09047 Selargius (CA), Italy 
              \and
              National Radio Astronomy Observatory, Socorro, NM 87801, USA
              \and 
              Naval Research Laboratory Remote Sensing Division, Code 7213 4555 Overlook Ave SW, Washington, DC 20375, USA
              }

   \date{Received MM DD, YY; accepted MM DD, YY}

 %
  \abstract
{Synchrotron radio emission 
is being detected from an increasing number of galaxy clusters. Spectral index images are a powerful tool
  to investigate the origin, nature, and connection of these sources with the dynamical
  state of the cluster.}
{The aim of this work is to investigate
    the spectral index distribution of the radio halo in the galaxy cluster A520, a
    complex system from an optical, radio, and X-ray point of view.}
{We present deep Very Large
Array observations in total intensity at 325 and 1400\, MHz. We
produced and analyzed spectral index images of the radio halo in this
frequency range at a resolution of 39$^{\prime\prime}$ and
60$^{\prime\prime}$ and looked for possible correlations with the
thermal properties of the cluster.}
{We find an integrated radio halo
spectral index $\alpha_{\rm 325}^{1400}\sim1.12$.  No strong radial
steepening is present and the spectral index distribution is
intrinsically complex with fluctuations only partially due to
measurement errors. The radio halo integrated spectral index and the cluster
temperature follow the global trend observed in other galaxy
clusters although a strong point-to-point correlation between the spectral index and the thermal gas temperature 
has not been observed. }
{The complex morphology in the spectral index image of the radio halo in
A520 is in agreement with the primary models for radio halo
formation. The flatness of the radial profile suggests that the merger is still ongoing and is uniformly and continuously (re-) accelerating the population of relativistic electrons responsible of the radio emission even at large ($\sim1$\,Mpc) distances from the cluster center.}
 

   \maketitle

\begin{table*}
\caption{Details of the VLA observations of Abell 520.}             
\label{observation_details}      
\centering          
\begin{tabular}{c c c c c c c c}     
\hline\hline       
RA     &  DEC    & $\nu$ & Bandwidth & Config. & Date & Duration & Project  \\
(J2000) & (J2000) & (MHz) &  (MHz)  & ~ & ~ & (Hours) & ~ \\
\hline
04:54:09.30  &     +02:55:21.00  & 328, 323    &6.25     & B &15,16-Apr-05  &5.5  &AC0776  \\
04:54:09.30  &     +02:55:21.00  & 328, 323    &6.25     & C &02,03-May-04  &6    &AC0706  \\
04:54:09.30  &     +02:55:21.00  & 1365, 1435      &50.0     & C &30-Aug-05  &5   &AC0776  \\
04:54:09.30  &     +02:55:21.00  & 1365, 1435      &50.0     & D &13-Aug-04  &7  &AC0706  \\
\hline    
 \multicolumn{8}{l}{\scriptsize Col. 1, Col. 2: Observation pointing; Col. 3: Observing frequency;}\\
\multicolumn{8}{l}{\scriptsize Col. 4: Bandwidth; Col. 5: VLA configuration; 
Col. 6: Observing dates; Col. 7: Total integration time;
Col. 8: Observation project.}\\         
\end{tabular}
\end{table*}

\begin{table*}
\caption{Information on total intensity images.}             
\label{image_details}      
\centering          
\begin{tabular}{c c c c} 
\hline\hline       
Array&$\nu$ & Beam                             & $\sigma(I)$ \\
~    &MHz   & $(^{\prime\prime})^{2}$ & mJy/beam    \\
\hline
B    &325   & 18 $\times$ 18                  &0.6           \\
C    &325   & 60 $\times$ 60                  &0.8             \\
B$+$C   &325   & 26 $\times$ 26                  &0.7           \\
C    &1400  & 16 $\times$ 16                  &0.025         \\
D    &1400  & 50 $\times$ 50                  &0.05          \\
C$+$D   &1400  & 26 $\times$ 26                  &0.03         \\
\hline    
 \multicolumn{4}{l}{\scriptsize Col. 1: VLA configuration;}\\
\multicolumn{4}{l}{\scriptsize Col. 2: Observing frequency;}\\
\multicolumn{4}{l}{\scriptsize Col. 3: Resolution of the observation;}\\      
\multicolumn{4}{l}{\scriptsize Col. 4: rms noise of the total intensity image.}\\         

\end{tabular}
\end{table*}

\section{Introduction}

The formation of massive galaxy clusters can be explained in the
context of the hierarchical scenario as due to the collision and
subsequent merger of small galaxy groups and subclusters. These
phenomena are the most energetic in the Universe since the Big Bang.
Energies as large as $\gtrsim$10$^{64}$\,ergs (e.g. Sarazin 2002) are
released in the form of shocks and turbulence that accelerate
particles and compress magnetic field, leading to large-scale diffuse
synchrotron sources associated with the intracluster medium (ICM) and
known as radio halos and relics.

Radio halos and relics are faint ($\sim\mu$Jy/arcsec$^2$ at 1.4\,GHz)
sources extended on Mpc scales, located respectively at the center and
in the outskirts of about 100 merging galaxy clusters (see Feretti et
al. 2012 for a recent review).  The study of radio halos and relics is
of paramount importance to shed light on the history and physical
properties of galaxy clusters, and to clarify the role of non-thermal
components associated to the ICM.  Their nature reveals very weak
large scale magnetic fields, with central strengths $\sim\mu$G,
fluctuation scales up to several hundreds of kpc (e.g. Vacca et
al. 2010), and rarefied, very energetic populations of relativistic
electrons spread over the cluster volume.

A key tool to investigate the shape of the relativistic electron
spectrum and the link between thermal and non-thermal properties in
galaxy clusters is the spectral index\footnote{Throughout this work we
  adopt the convention $S(\nu)\propto\nu^{-\alpha}$.} distribution of
halos and relics.  Spectral index images have been produced so far
for just a few radio halos (e.g. Coma Giovannini et al. 1993, A665 and
A2163 Feretti et al. 2004, A3562 Giacintucci et al. 2005, A2744 and
A2219 Orr\`u et al. 2007, A2255 Pizzo \& de Bruyn 2009, A2256 van
Weeren et al. 2009, Kale \& Dwarakanath 2010) revealing complex
distributions with flattening in the regions directly influenced by
ongoing mergers.  Investigations of the integrated spectrum of radio
halos have been performed as well, as in Coma
(Thierbach et al.  2003), A2319 (Feretti et al. 1997), A754 and A1914
(Bacchi et al. 2003), A3562 (Giacintucci et al. 2005), A2256
(Brentjens 2008, Kale \& Dwarakanath 2010), A697 (van Weeren et
al. 2011, but see also Venturi et al. 2013). 
In some cases a steepening at increasing frequencies has
been detected (e.g., Coma, A2319, A754, A3562) that could be due to 
the Sunyaev-Zel'dovich (SZ) effect (Ensslin et al. 2002, Pfrommer \& Ensslin 2004) and/or to strong energy losses in the relativistic electron population. In the case of the diffuse emission of the Coma cluster Brunetti et al. (2013) recently showed that the SZ decrement does not significantly impact on the shape of the radio halo spectrum.

Despite the improvements in the capabilities of present radio
interferometers and in the analysis procedures, a number of questions
concerning radio halo origin and evolution are still open. A mechanism
of (re-) acceleration of relativistic electrons (primary models,
e.g. Brunetti et al. 2009) or \emph{in situ} generation (secondary models,
e.g. Ensslin et al. 2011) is required to explain their emission
over Mpc scales.  Observational properties of radio halos, e.g. the
complex radio-halo morphology and spectral index distribution, high
frequency cut-off in the radio halo spectrum, and spectral radial
steepening support primary model predictions. 
Nevertheless, Ensslin et al. (2011) show that the radio halo
formation could be explained also by means of a high central energy
density concentration of cosmic rays that should be typical of cluster
undergoing major mergers and strongly turbulent.  We have to note that
at the moment neither the primary nor the secondary models allow an
exhaustive description of the observational properties of radio
halos. It seems more likely that relativistic electrons continuously
generated by hadronic collisions and present in all galaxy clusters
coexist with a population of relativistic electrons (re-) accelerated
through magneto-hydro-dynamical turbulence due to cluster mergers (see
Cassano 2009, and references therein). The global picture has become
even more complex because of the recent discovery of radio halos in
low X-ray luminosity clusters (see Giovannini et al. 2011 and
references therein) that opens new questions concerning the radio halo
formation processes. Detailed analysis of the radio halo emission at
different frequencies are essential to shed light on these aspects.

In this paper we present the study of the spectral index of the radio
halo in the galaxy cluster Abell 520 between 325 and 1400\,MHz.  In
\S\,\ref{Abell520} a description of the present radio, X-ray, and
optical knowledge on A520 is given. In \S\,\ref{Radio observations and
  data reduction} the radio observations and data reduction procedures are
presented. In \S\,\ref{Spectral index analysis} the results about the
spectral index of the radio halo and the radio galaxies in A520 are
shown. In \S\,\ref{Discussion} we discuss the results, while in
\S\,\ref{Conclusions} we draw our conclusions.

Throughout this paper we adopted a $\Lambda$CDM cosmology with
$H_0=71$ km s$^{-1}$ Mpc$^{-1}$, $\Omega_{m}=0.27$, and
$\Omega_{\Lambda}=0.73$. At the distance of A520 (z=0.199, Struble \&
Rood 1999), 1\arcsec\, corresponds to 3.25\,kpc.

\section{The cluster of galaxies Abell 520}
\label{Abell520}
\begin{figure*}[ht]
  \centering
  \includegraphics[width=13cm, angle=0]{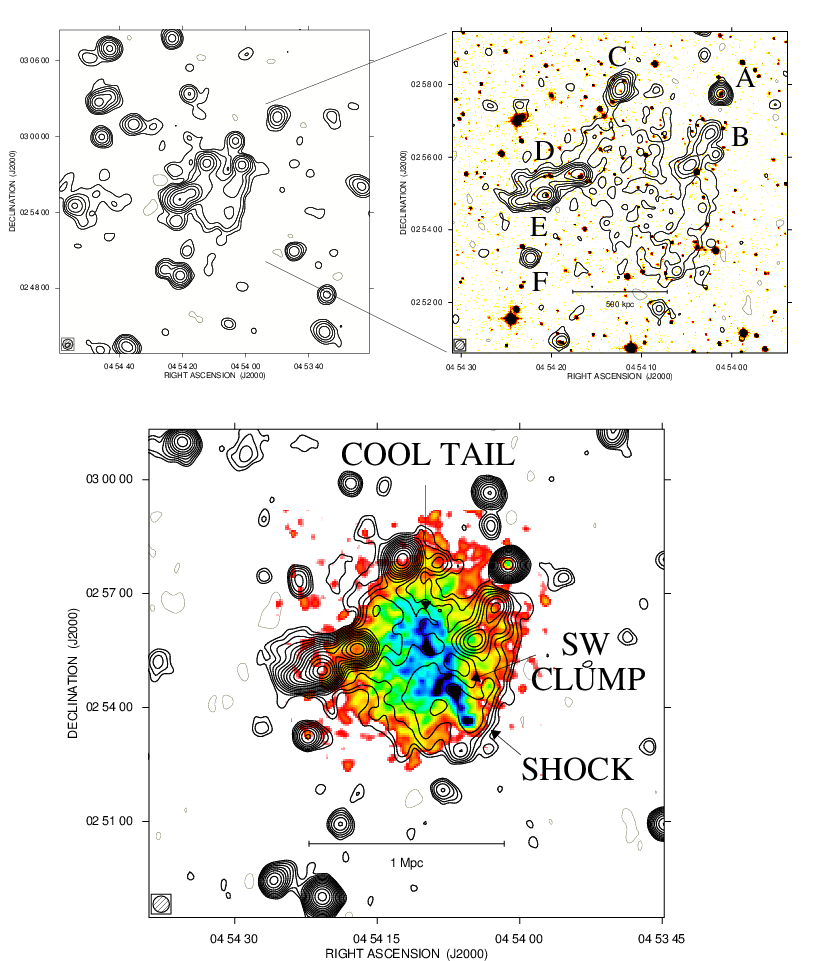}
  \caption{\emph{Top left panel:} total intensity radio contours at
    1400\,MHz (VLA in D configuration) with an FWHM of
    $50^{\prime\prime}\times50^{\prime\prime}$ . The contour levels
    are -0.15\,mJy/beam, 0.15\,mJy/beam, and the rest spaced by a
    factor 2. \emph{Top right panel:} total intensity radio contours
    at 1400\,MHz (VLA data in C configuration) with an FWHM of
    $16^{\prime\prime}\times16^{\prime\prime}$ . The contour levels
    are -75\,$\mu$Jy/beam, 75\,$\mu$Jy/beam, and the rest spaced by a
    factor 2. The contours of the radio intensity are overlaid on the
    red plate of the Sloan Digital Sky Survey. \emph{Bottom panel:}
    total intensity radio contours at 1400\,MHz (combining VLA data in
    C and D configuration) with an FWHM of
    $26^{\prime\prime}\times26^{\prime\prime}$ . The contour levels
    are -90\,$\mu$Jy/beam, 90\,$\mu$Jy/beam, and the rest spaced by a
    factor $\sqrt{2}$. The contours of the radio intensity are
    overlaid on the ACIS-I \emph{Chandra} X-ray adaptively smooth brightness image in the
    0.8--4\,keV band (Govoni et al. 2004) in colors.  }
  \label{A520_1.4_all}
\end{figure*}

Abell 520 is a complex and interesting galaxy cluster that has been
observed in the X-ray, optical, and radio domain.  By analysing NRAO
Very Large Array (VLA) Sky Survey (NVSS, Condon et al. 1998) images,
Giovannini et al. (1999) found for the first time hints of diffuse
large scale emission in this cluster, later confirmed by the work of
Govoni et al. (2001). Govoni et al. (2001) reveal a wide radio halo
with
a largest linear size of 1.4\,Mpc, elongated in the NE-SW
direction, as the central X-ray emission. Indeed, the ROSAT low
resolution X-ray observations indicate that the X-ray emission
consists of an inner elongation in the NE-SW direction and of an outer
emission in the SE-NW direction.
High resolution \emph{Chandra} X-ray observations by Govoni et
al. (2004) indicate that this system is undergoing a strong merger
event in the NE-SW direction. They detect a dense, cool gas clump in
the South-West of the cluster, probably coming from North-East, and
followed by a cool tail characterized by a strip of hot gas where the
radio halo emission seems to originate from (see the bottom panel of 
Fig.\,\ref{A520_1.4_all}).

Merger shocks have been firmly detected only in a
few clusters from jumps in X-ray surface brightness and in temperature. Some examples
are in A3667 (Vikhlinin et al. 2001, Sarazin et al. 2013) and in Coma (Neumann et al. 2001, Akamatsu et al. 2013, Ogrean \& Brueggen 2013). Another example is given by A2146, where Russel et al. (2010) found a bow shock from a sharp discontinuity in the temperature and gas density. 
Deep \emph{Chandra} observations by Markevitch et
al. (2005) of A520 indicated the presence of
a bow shock, previously suggested by Govoni et al.  (2004), in front of the dense clump, 
coincident with the SW edge of the radio halo. 
This shock is characterized
by a Mach number $M=2.1_{-0.3}^{+0.4}$ and is the second
unambiguously recognized supersonic merger shock front in a cluster
that shows a sharp gas density edge and a clear temperature jump (the
first one is the Bullet cluster, Markevitch et al. 2002). 
Markevitch et al. (2005)
state that the electrons responsible for the edge of the radio halo
emission in A520 may be accelerated by this shock and, in this hypothesis,
they expect a slope of the radio halo spectrum $\alpha\simeq1.2$ in
correspondence of the edge. The spectrum should quickly steepen as a function
of the distance from the edge in the post-shocked region, since no turbulence or other kind of acceleration mechanism are supposed to be present in between the shock and the cool tail in the NE. This trend has been observed in elongated relics, as for example in A1240 (Bonafede et al. 2009b) and in CIZAJ2242.8+5301 (van Weeren et al. 2010), where the spectral index steepens toward the cluster center. On the
contrary if the efficiency of the shock is not enough to accelerate the
electrons, the radio edge may be caused by the compression of the
magnetic field. In this scenario, they expect to detect a 10-20 times
fainter radio emission with the same spectrum in front of the shock where the
pre-compressed relativistic particles and
magnetic fields exist.

Optical investigations by Proust et al. (2000) indicate as well a
complex system hosting a concentration of galaxies at the center of
the cluster with two extensions. They suggest that this cluster is a
dynamically young system and these extensions are related to clumps of
galaxies colliding on a dark matter gravitational well, located at the
center of the X-ray emission region.  This insight has been confirmed
by the analysis of Jee et al. (2012).  By means of a high resolution
weak lensing studies carried out with Hubble Space Telescope data,
they map a filamentary structure of 1.5\,Mpc in size elongated in the
NE-SW direction and confirm the presence of a dark matter core with a
significance larger than 10$\sigma$, whose presence has been already
observed by Mahdavi et al. (2007) at 4$\sigma$ level. In contrast with
observations in other merging galaxy clusters (e.g., the Bullet
cluster, Clowe et al. 2006) and with the collisionless dark matter
scenario, the dark matter core is at the same location as the X-ray
luminosity distribution peak but not as the bright cluster galaxies.
Results from Girardi et al. (2008) from TNG and INT observations
combined with data from literature and from the CNOC team (Carlberg et al. 1996, Yee et al. 1996) suggest
that the cluster formation is actually taking place at the crossing of
three filaments: one in the NE-SW direction, one in the EW direction,
and one almost aligned with the line of sight.  The peak in the dark
matter distribution could be therefore the consequence of projection
effects.

By analyzing new Hubble Space Telescope data, Clowe et al. (2012) do
not detect the dark matter core and, after subtraction of the X-ray plasma
mass, they find a good agreement between the mass distribution in this cluster and the luminosity distribution of the cluster galaxies, and between weak lensing mass measurements and the morphology of the core galaxy-filled structure.
Moreover, they find that the mass within a sphere encompassing a mean
overdensity of 200 is $M_{\rm 200}=(9.1\pm1.9)\times
10^{14}M_{\odot}$, in agreement with previous works.

\section{Radio observations and data reduction}
\label{Radio observations and data reduction}
We present archival observations (Projects AC0776 and AC0706) 
of A520 at 325 and 1400\,MHz obtained
using the Very Large Array (VLA) respectively in spectral line and in
continuum mode. The observations at 325\,MHz have been obtained in B
and C configurations, while those at 1400\,MHz are in C and D
configurations. The details of the observations are summarized in
Table\,\ref{observation_details}.  The data reduction has been
performed following standard procedures using the NRAO's Astronomical
Image Processing System (AIPS) package. Total intensity images were
produced with the multi-scale clean technique (see e.g.  Greisen et
al. 2009), an extension of the classical clean algorithm implemented
in the task IMAGR. The fluxes are expressed according to Perley - Taylor 1999.2 scale \footnote{See the \emph{VLA Calibrator Manual} \protect\url{http://www.vla.nrao.edu/astro/calib/manual/baars.html}.}.

At 1400\,MHz the radio source 0542+498=3C147 was used as primary flux
density calibrator while the nearby source 0503+020 has been used as
complex gain calibrator. Radio interferences were carefully excised.
The two IFs were averaged to obtain the surface brightness image and
several cycles of self-calibration and CLEAN were applied.

The 325\,MHz data were obtained with a total bandwidth of 6.25\,{\rm
  MHz}, subdivided in 16 channels with a resolution of 390\,{\rm
  kHz} each. We excluded the edge channels and averaged over channels, resulting in 5 channels with a resolution of $\sim$1\,{\rm
  MHz} each. The radio source 0542+498=3C147 was used as primary flux
density and bandpass calibrator while the nearby source 0521+166=3C138
has been used as complex gain calibrator. Radio interferences were
carefully excised with a flag of about 30\% of the data.  To obtain
the surface brightness image we averaged the five channels of both IFs
together in the gridding process using IMAGR.  Since VLA images over a
wide-field-of-view suffer from distortions due to the non-coplanarity of the array (Perley 1999),
we cover the central $\sim4^{\circ}$ with several small overlapping
facets. Strong sources outside this area can affect the final image
because their secondary lobes can be present in our field of view at
this frequency.  To include in the cleaning process these sources, we
looked for sources with a flux larger than 0.5\,Jy at 1400\,MHz in a
region of about $\sim 6^{\circ}$ in radius, by using the NVSS
catalog. These sources have been included in the cycles of
self-calibration and CLEAN applied to remove residual phase
variations. After self-calibration, to better image the diffuse
emission at the center of the cluster, discrete sources at a linear
distance larger than $\sim$1.5\,Mpc have been subtracted.

Moreover, for each frequency, data corresponding to different
configurations of the VLA have been combined aiming at producing
images with better uv-coverage and sensitivity but still good angular
resolution.  The sensitivities of the final images are summarized in
Table\,\ref{image_details}. We note that our sensitivity at 325\,MHz
in C configuration $\sim0.8$\,mJy/beam is slightly higher than the
theoretical noise for the bandwidth and the duration of the
observations ($\sim$0.5\,mJy/beam, of the same order of the confusion limit at
this frequency $\sim$0.44\,mJy/beam), while at 1400\,MHz our
sensitivity in D configuration $\sim$50\,$\mu$Jy/beam is limited by
the confusion.

\begin{figure}[h]
  \centering
  \includegraphics[width=7cm, angle=0]{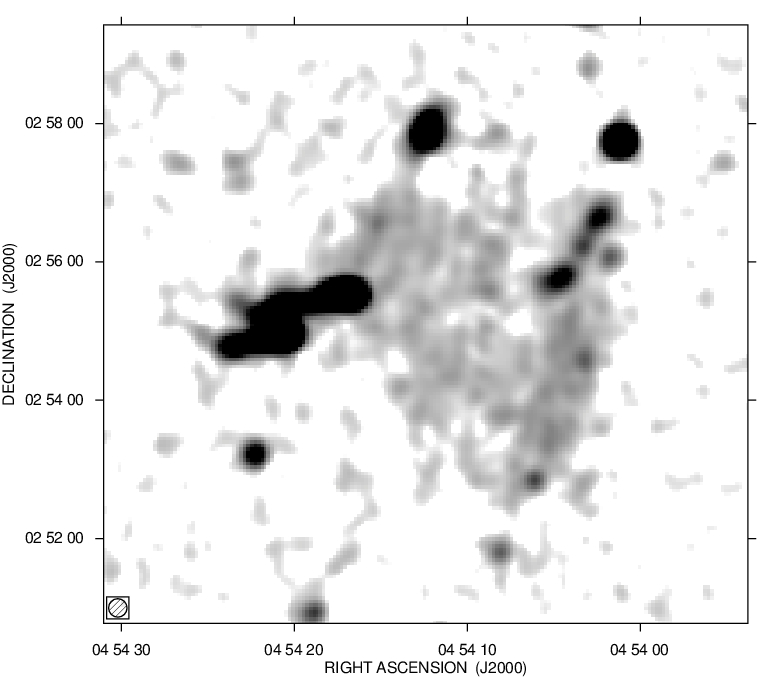}
  \caption{Total intensity radio brightness
    at 1400\,MHz (VLA data in C configuration) with an FWHM of
    $16^{\prime\prime}\times16^{\prime\prime}$.}
  \label{A520_C_grey}
\end{figure}

The 1400\,MHz total intensity images of the radio emission in A520 at
different resolutions are shown in Fig.\,\ref{A520_1.4_all}.  The top
left panel shows the image in D configuration where the full extent of
the radio halo is visible. After correcting for the primary beam and
masking the radio galaxies, from this image we measure a flux of the
radio halo at 1400\,MHz $S_{\rm 1400\,MHz}=(16.7\pm0.6)$\,mJy. 
Here and in the following, the total error is the quadratic sum of the statistical error and of the systematic error. The statistical error includes calibration, while the systematic error is related with flux-scale uncertainty that has been assumed to be 3\% of the measured flux in order to take into account also the bootstrap uncertainty.
We note that this value of the flux differs from the 34\,mJy measured by Govoni et al. (2001) probably because of differences in the subtraction of the point sources. From now on the value of ($16.7\pm0.6$)\,mJy should be considered as the reference. 
Subtracting the flux of discrete sources measured with C configuration data from the total flux over the region covered by the diffuse emission measured with D configuration gives an estimation consistent with this value within the errors. A higher
resolution image obtained from the data in C configuration is shown in
the top right panel, but see also Fig.\,\ref{A520_C_grey} for a greyscale visualization. The radio contours are superposed to an optical
image from the \emph{Sloan Digital Sky Survey} red plate. Radio
galaxies at the boundary of the diffuse emission can be easily
identified. Some of them present an optical counterpart, i.e. the
source A and the two narrow angle tailed sources D and E.  In the
bottom panel a combination of the datasets in D and in C configuration
is shown. The radio isocontours are overlaid on an ACIS-I
\emph{Chandra} X-ray image in the 0.8--4\,keV band (Govoni et al. 2004).  
The cluster appears characterized by substructures on scales $\gtrsim 80$\,kpc both at the radio and at the X-ray wavelengths.  
In the cluster center as well as in the
outskirts, where the bow shock has been detected, peaks and
substructures in the X-ray and in the radio image have the same
spatial location, indicating a deep link between thermal and
non-thermal properties in this system.

The total intensity images of the galaxy cluster at 325\,MHz are shown
in Fig.\,\ref{A520_327_B_C}. On the top left panel, the low resolution
image is presented. The radio halo appears extended as in the low
resolution image at 1400\,MHz (top left panel in
Fig.\,\ref{A520_1.4_all}), while the emission of some discrete sources
observed at 1400\,MHz is below the noise level in this image, giving
an upper limit for their spectral index (see \S\,\ref{Spectral index
  radio galaxies}). After correcting for the primary beam and masking
radio galaxies, from this image we measure a flux density for the
radio halo of $S_{\rm 325\,MHz}=(85\pm5)$\,mJy.  On the top right
panel, a high resolution image at 325\,MHz is shown where the radio
halo emission is completely resolved.  The combination of the B and C
configuration observations is shown in the bottom panel.

The integrated spectral index evaluated from the above fluxes is 
$\alpha_{325}^{1400}=1.12\pm0.05$.
The spectral index of discrete radio sources are presented in \S\,\ref{Spectral index radio galaxies}.

\section{Spectral index images}
\label{Spectral index analysis}
To produce spectral index images of the radio emission at the center
of the cluster we compared total intensity images at 325 and 1400\,MHz
using images produced with the same uv-coverage, resolution and pixel size, and the Dan Briggs AIPS robustness parameter set to zero.  
As discussed in Orr{\`u} et al. (2007), considering only those pixels with a
signal larger than 3$\sigma$ at both frequencies implies a bias in the
measurement of the spectral index, since regions with a flat spectrum
can not be investigated. The faintest regions of the radio halo at
1400\,MHz are above the noise at 325\,MHz only if the radio halo
spectral index is steeper than $\sim$2.  To minimize this effect, in the analysis of the radial profile of the spectral index we directly compared the total intensity images at
325 and 1400\,MHz without applying any cut in brightness, unless
otherwise specified.

\subsection{Radio halo}
\label{Radio halo spectral index analysis}
\begin{figure*}[ht]
  \centering
  \includegraphics[width=15cm, angle=0]{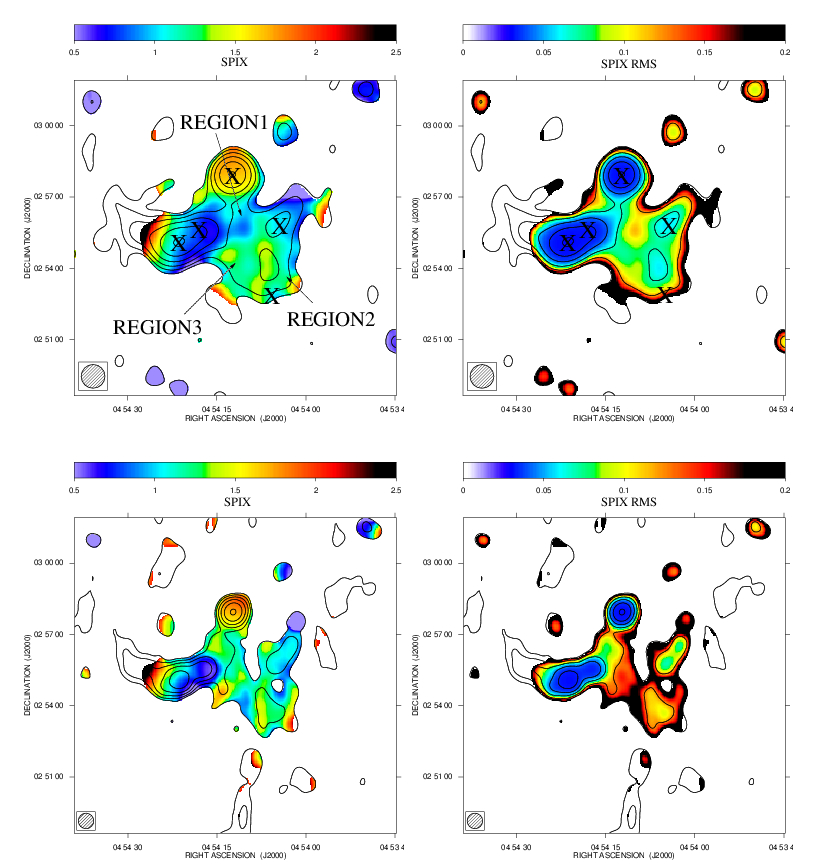}
  \caption{Radio contours at 325\,MHz overlaid on the spectral index
    image (left panels) and on spectral index error image (right
    panels) of A520 between 325 and 1400\,MHz with a resolution of
    $60^{\prime\prime}\times60^{\prime\prime}$ (top panels) and
    $39^{\prime\prime}\times39^{\prime\prime}$ (bottom panels). We
    blanked pixels with brightness below 3$\sigma$ at 325 or
    1400\,MHz. The crosses in the low resolution image represent the radio galaxies embedded in the diffuse emission. }
      \label{Spix_all}
\end{figure*}

In Fig.\,\ref{Spix_all} the image of the spectral index of the radio halo and the radio galaxies embedded in the diffuse emission and the image of the 1$\sigma$ uncertainty evaluated pixel by pixel are shown at 60$^{\prime\prime}$ ($\sim$200\,kpc), top panels, and at 39$^{\prime\prime}$ ($\sim$120\,kpc), bottom panels.

The low resolution image allows better recovery of the faint structures
in the North and in the South-East, where flatter spectral index
values can be observed.  We evaluated the spectral index in the
regions indicated in Fig.\,\ref{Spix_all} (top left panel) by using
boxes 1.5 times the beam in size. The North (REGION1 in Fig.\,\ref{Spix_all}) 
shows an average
spectral index of $\langle\alpha\rangle=1.03$, the South-West
(REGION2) of $\langle\alpha\rangle=1.22$ and the South-East (REGION3)
of $\langle\alpha\rangle=1.23$. 
These values are consistent with those evaluated directly from the integrated fluxes at the two frequencies.

In Fig.\,\ref{A520_spix_radialprofile_beam} the azimuthally averaged
brightness profile at 325 and 1400\,MHz (top panel) and the consequent
spectral index (bottom panel) are shown. Since we are interested on
the average radial behavior of the spectral index, we measured the
surface brightness from the 60$^{\prime\prime}$ resolution
images. Each data point with its 1-$\sigma$ uncertainty represents the
average brightness in concentric annuli of half-beam-width centered on
the X-ray peak\footnote{As X-ray peak we adopted the value give by
  Govoni et al. (2001), RA (J2000) =$04^{\rm h}54^{\rm m}10.6^{\rm s}$
  and DEC (J2000)=$+02^{\circ}55^{\prime}20^{\prime\prime}$.}, as
shown in the inset. The arrows represent the upper limits at
3-$\sigma$.  Discrete sources have been masked out and excluded from
the statistics. Their flux and spectral index are given in
\S\,\ref{Spectral index radio galaxies}.  The spectral index profile
of the radio halo appears rather flat.

\begin{figure*}[ht]
  \centering
  \includegraphics[width=13cm, angle=0]{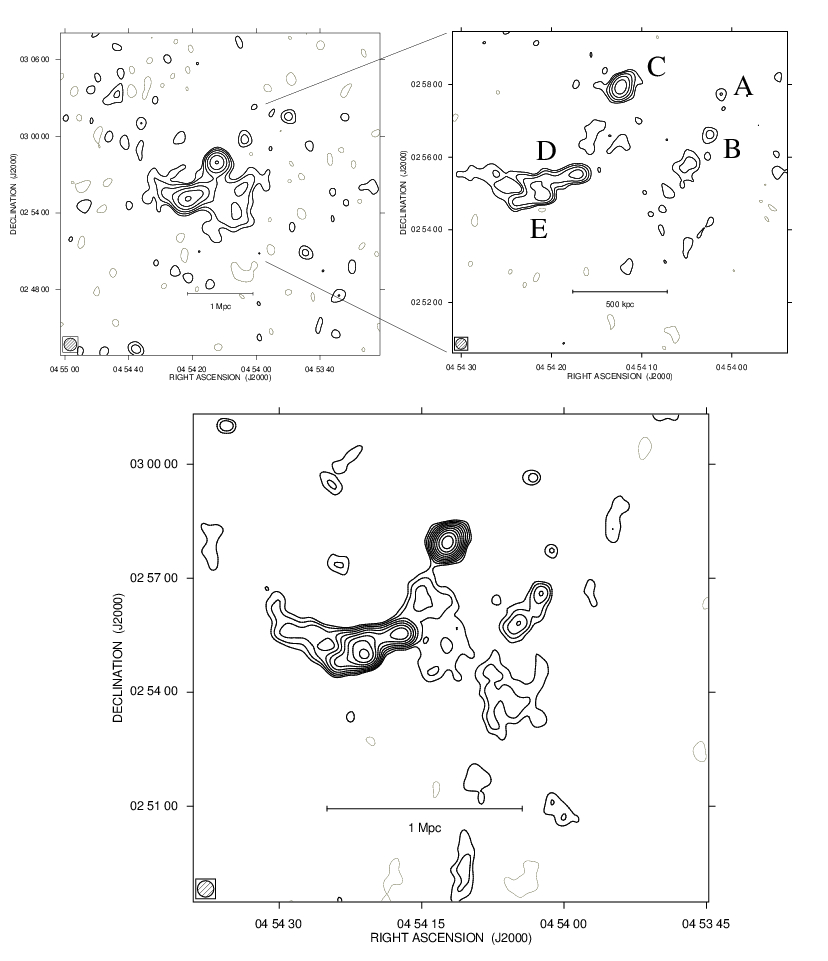}
  \caption{\emph{Top left panel:} total intensity radio contours at
    325\,MHz (VLA in C configuration) with an FWHM of
    $60^{\prime\prime}\times60^{\prime\prime}$ . The contour levels
    are -2.4\,mJy/beam, 2.4\,mJy/beam and the rest spaced by a factor
    2. \emph{Top right panel:} total intensity radio contours at
    325\,MHz (VLA data in B configuration) with an FWHM of
    $18^{\prime\prime}\times18^{\prime\prime}$ . The contour levels
    are -1.8\,mJy/beam, 1.8\,mJy/beam and the rest spaced by a factor
    2. \emph{Bottom panel:} total intensity radio contours at 325\,MHz
    (combining VLA data in B and C configuration) with an FWHM of
    $26^{\prime\prime}\times26^{\prime\prime}$ . The contour levels
    are -2.1\,mJy/beam, 2.1\,mJy/beam and the rest spaced by a factor
    $\sqrt{2}$. }
              \label{A520_327_B_C}
    \end{figure*}

\begin{figure}[h]
   \centering
  \includegraphics[width=9cm, angle=0]{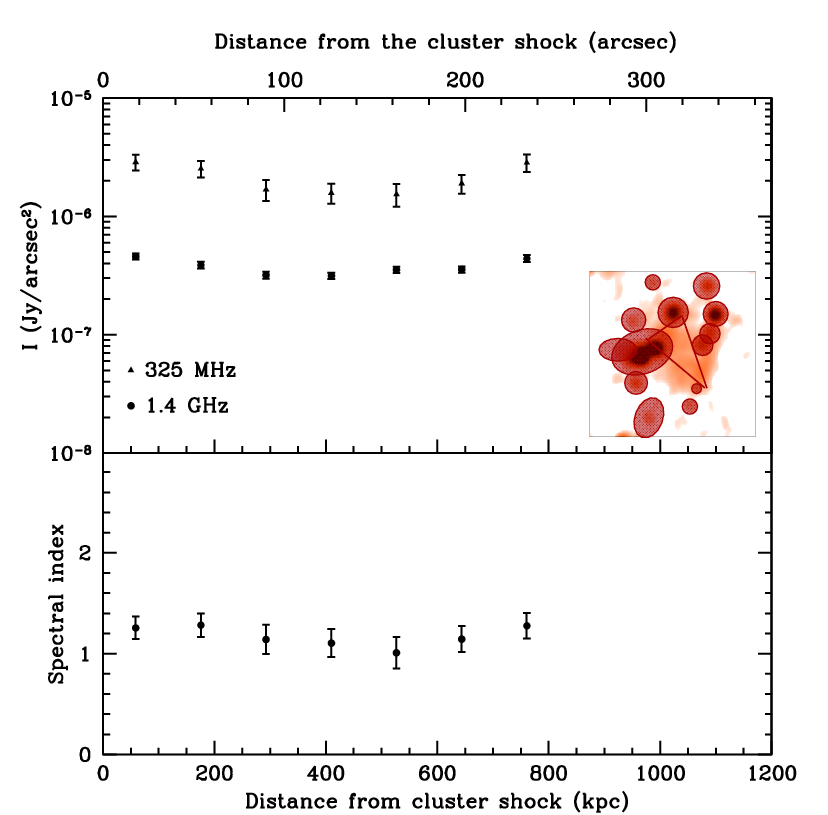}
      \caption{\emph{Top:}  azimuthally averaged brightness of the
        radio halo at 325 (triangles) and 1400\,MHz (circles) in a sector of concentric annuli versus
        the distance from the shock peak at a resolution of
        39$^{\prime\prime}$.  The sector starts from the
        South-West and crosses the cluster towards 
        the North-East, see the inset. \emph{Bottom:} spectral index along the sector.}
              \label{A520_spix_grid}
    \end{figure}

In Fig.\,\ref{A520_spix_grid}  
the radio brightness at 325 and 1400\,MHz (top panel) and the consequent
spectral index (bottom panel) along a sector of the diffuse emission are shown.
Each data point with its 1-$\sigma$ uncertainty represents the
average brightness in a sector of concentric annuli of beam-width centered on
the shock peak and crossing the direction of propagation of the shock and the NE periphery along the cool tail, see the inset panel. To be more sensitive to
small scale variations we used the total intensity images at
39$^{\prime\prime}$.  
The brightness of the radio halo keeps flat both at 325 and at
1400\,MHz, with a consequent flat spectral index.

\subsubsection{Fluctuations in the radio halo spectral index image}
\label{Fluctuactions in the radio halo spectral index image}

In Table\,\ref{statistics} the statistics of the spectral index and
the spectral index error images of the radio halo both at
39$^{\prime\prime}$ and at 60$^{\prime\prime}$ (Fig.\,\ref{Spix_all}) are summarized.  The mean values of the spectral index and of its error 
are in good agreement at the two resolutions. 
\begin{table}
\caption{Statistics of the spectral index and the spectral index error images.}
\label{statistics}      
\centering          
\begin{tabular}{c c c}     
\hline\hline       
Beam & $\langle\alpha\rangle$ &$\langle Err_{\alpha}\rangle$\\
$(^{\prime\prime})^2$&&\\
\hline
39$\times$39&1.25&0.16\\
60$\times$60&1.21&0.12\\
\hline    
\end{tabular}
\end{table}

\begin{figure}[h!]
  \centering
  \includegraphics[width=8cm, angle=0]{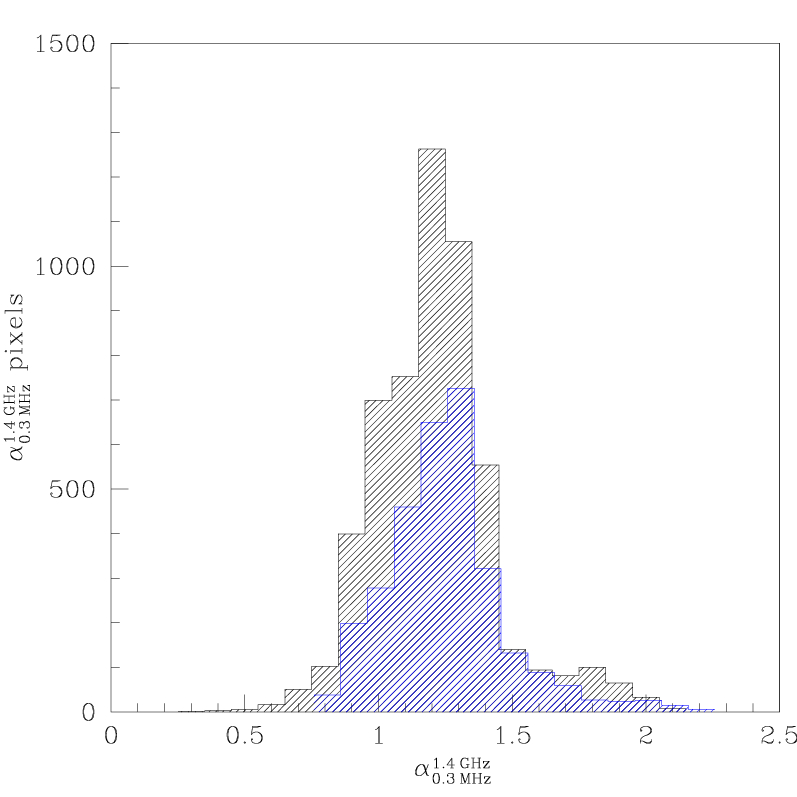}
  \caption{Histogram of the spectral index values from the image at
    60$^{\prime\prime}$ (black) and 39$^{\prime\prime}$ (blue)
    resolution. These distributions have a mean value and a dispersion
    $\langle\alpha_{60^{\prime\prime}}\rangle=1.21$ and
    $\sigma_{\alpha_{60^{\prime\prime}}}=0.23$ at 60$^{\prime\prime}$
    and $\langle\alpha_{39^{\prime\prime}}\rangle=1.25$ and
    $\sigma_{\alpha_{39^{\prime\prime}}}=0.22$ at
    39$^{\prime\prime}$.}
      \label{Spix_histo}
\end{figure}

The spectral index image of the radio halo appears clumpy, with
fluctuations both at high and at low resolution.  
To investigate the
nature of these fluctuations we studied the distribution of the
spectral index values. We extracted the information pixel by pixel from the spectral index and the spectral index error image
after masking the radio galaxies (Fig.\,\ref{Spix_histo}) and we calculated the mean and the sigma over these values. 
The blue
histogram refers to the 39$^{\prime\prime}$ image, while the black one
to the 60$^{\prime\prime}$ image.  The two distributions have a mean
value $\langle\alpha_{60^{\prime\prime}}\rangle=1.21$ and dispersion
$\sigma_{\alpha_{60^{\prime\prime}}}=0.23$ at 60$^{\prime\prime}$, and
$\langle\alpha_{39^{\prime\prime}}\rangle=1.25$ and
$\sigma_{\alpha_{39^{\prime\prime}}}=0.22$ at 39$^{\prime\prime}$.
The distributions are asymmetric respect to the mean value both at high and at low resolution. Indeed, they miss the flatter spectral index values, because of the cut in total intensity applied to produce the two spectral index images.

The error image of the spectral index has been produced evaluating the
uncertainty on a pixel basis, therefore if the patchy structure of the
spectral index image is due to measurement errors we expect that the
mean value of the error image and the dispersion of the spectral index
distribution to be comparable. When compared with the mean value of the
spectral index error image, the dispersion of the distributions in  Fig.\,\ref{Spix_histo} is slightly larger
both at high and at low resolution.

We conclude that the measurement errors significantly contribute to the observed
fluctuations in the spectral index distribution. 
Moreover, the spatial frequency coverage could cause instrumental noise responsible of fluctuations on spatial scales that are about the resolution size.
Nevertheless, a certain 
degree of intrinsic complexity seems to be present.

\subsubsection{Radio halo spectral index and X-ray properties of the cluster}
\label{Radio halo spectral index and X-ray properties of the cluster}

A520 is one of the first clusters where a comparison between the shape
of the radio and X-ray brightness profile has been performed (Govoni et al. 2001). They found that the two azimuthally averaged
profiles do not show a close similarity as has been observed in other
clusters (e.g. Coma, A2255, A2319), the radio halo being elongated
while the X-ray emission quite extended and symmetric.

\begin{figure}[ht]
  \centering
  \includegraphics[width=7cm, angle=0]{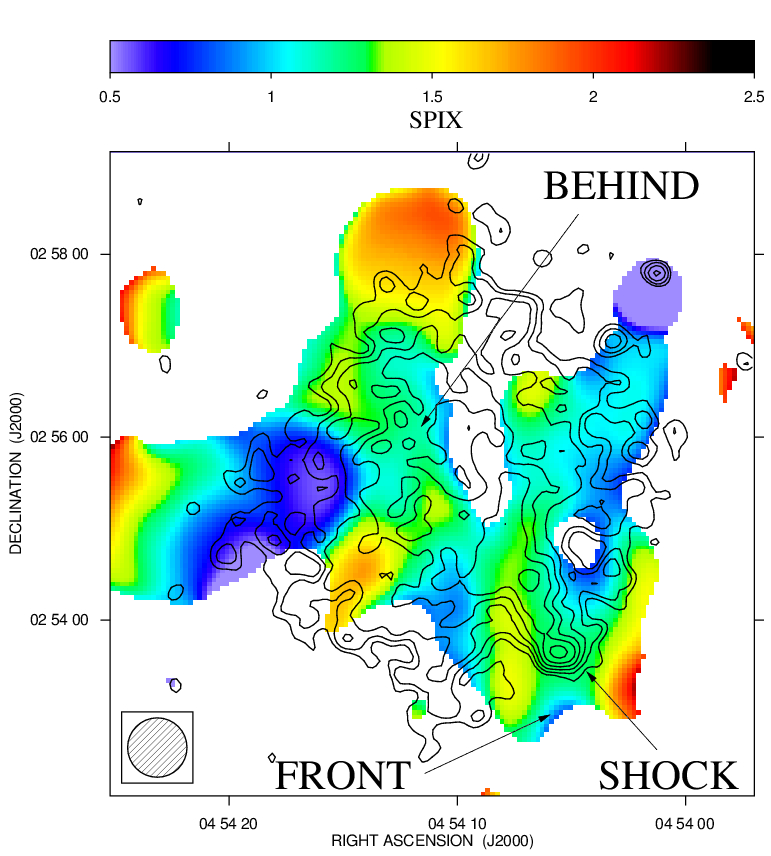}
  \caption{X-ray contours from the 0.8--4 keV X-ray image overlaid on
    the spectral index image of the radio halo at
    39$^{\prime\prime}$. The contours are spaced by a factor of
    $\sqrt{2}$.}
  \label{Spix_zoom}
\end{figure}

To infer some information about a possible link between the radio halo
emission and the thermal properties of the hosting cluster we compared
the spectral index with the X-ray brightness and the thermal gas
temperature images.  In Fig.\,\ref{Spix_zoom} the radio halo spectral
index image at 39$^{\prime\prime}$ superimposed with the contours from
the \emph{Chandra} X-ray brightness image in the 0.8--4\,keV band is shown.
The bright region coincident with the shock detected in the X-ray
image is characterized by an average spectral index $\sim$1.25 that
flattens to values $\sim$1--1.1 behind (in the North-East) and in
front (in the South) of the shock.
 
\begin{figure}[ht]
   \centering
   \includegraphics[width=7cm, angle=0]{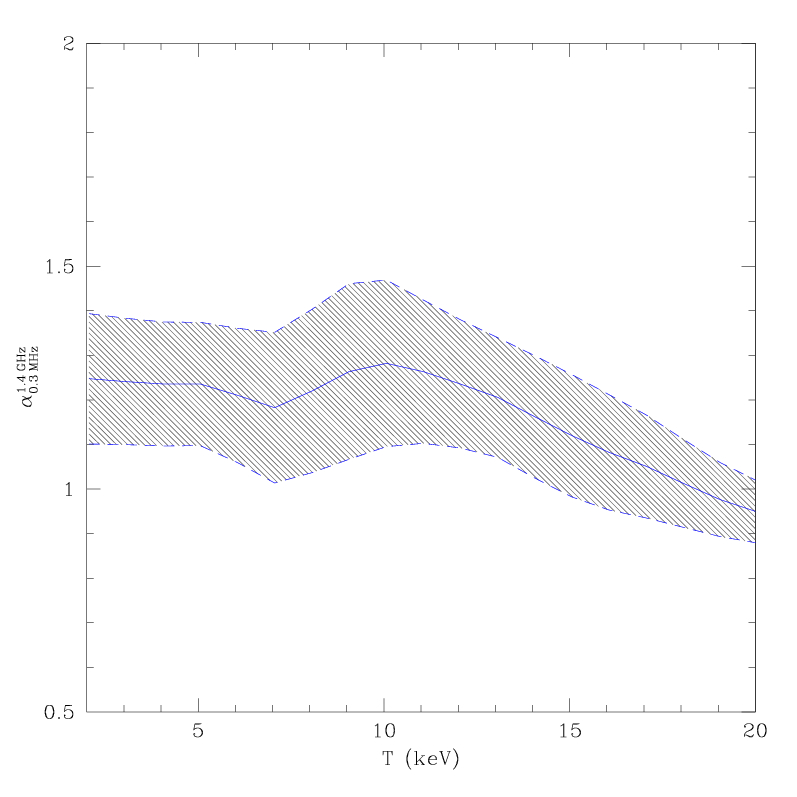}
  \caption{Exponential smooth of the spectral index $\alpha$ between
    325 and 1400\,MHz versus thermal gas temperature $T$. We adopted a smoothing scale of 1\,keV.}
              \label{Temperature}
    \end{figure}

A point-to-point correlation between the radio-halo spectral index
and the thermal-gas temperature has been observed for the first time
in the cluster in A2744 (Orr{\`u} et al. 2007), indicating that
regions of the radio halo with a flat spectrum appear characterized by
higher temperatures. We looked for the existence of such a correlation
for the radio halo in A520.  
To be more sensitive to point-to-point
fluctuations in the spectral index, we used the image at $39^{\prime\prime}$. For the temperature we used the image from Govoni et al (2004). In
Fig.\,\ref{Temperature} the exponential smoothing of the spectral
index between 325 and 1400\,MHz versus thermal gas temperature is
shown.  The data have been extracted pixel by pixel and then exponentially smoothed, i.e. a weighted moving average with exponential weights has been performed, with a smoothing scale of 1\,keV. Discrete sources have been masked out and excluded from the statistics. The shaded region describes the 1$\sigma$ uncertainty.   
In the coldest regions ($\sim$2\,keV), the
spectral index is between 1.1 and 1.4. Moving to higher 
temperatures, e.g. in the South and in the North-East of the cluster, the spectral index decreases to $\sim$0.9 with a scatter of
$\sim$0.05.

\begin{figure}[h]
   \centering
  \includegraphics[width=9cm, angle=0]{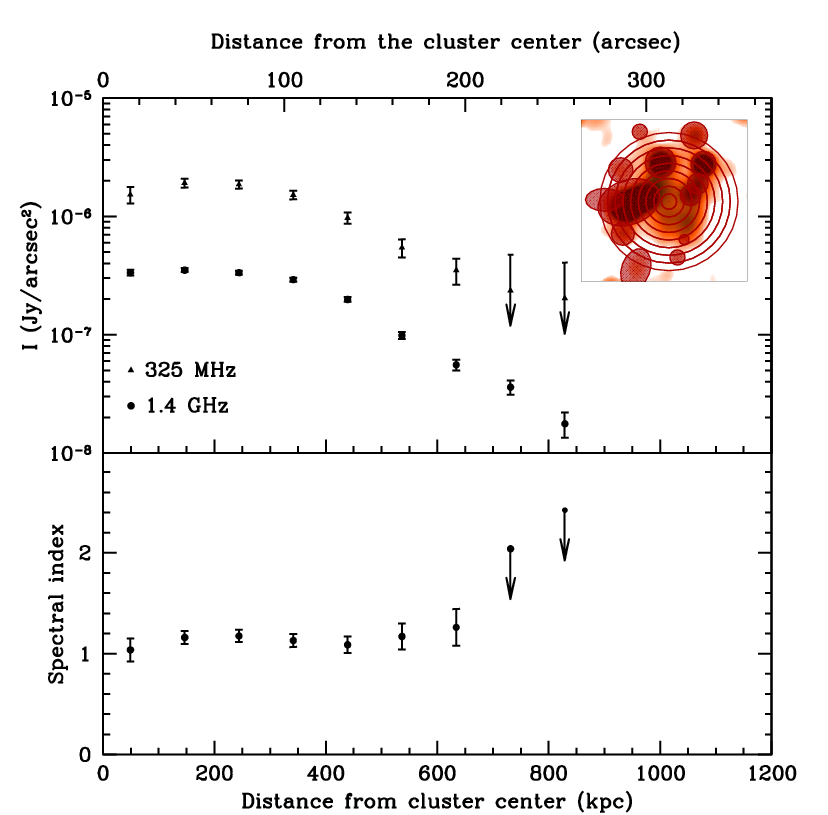}
      \caption{\emph{Top:} azimuthally averaged brightness of the
        radio halo at 325 (triangles) and 1400\,MHz (circles) versus
        the distance from the cluster center at a resolution of
        60$^{\prime\prime}$. Arrows represent 3-$\sigma$ upper
        limits. In the inset the annuli used to calculate the
        azimuthally averaged brightness are shown. Masks have been used
        to exclude discrete sources from the
        statistics. \emph{Bottom:} spectral index radial profile.}
              \label{A520_spix_radialprofile_beam}
    \end{figure}

If a global temperature of the thermal gas in A520 T=(7.1$\pm$0.7)\,keV is considered according to Govoni et al. (2004), the values
of spectral index and temperature observed in A520 appear to follow
the trend found for the other radio halos by Giovannini
et al. (2009) and Feretti et al. (2012).


\subsection{Radio galaxies}
\label{Spectral index radio galaxies}
A dominant radio galaxy can not be identified at the center of the
cluster but several powerful sources are present in the outskirts of
the diffuse radio halo emission.  These radio sources are labeled in
Fig.\,\ref{A520_1.4_all} and in Fig.\,\ref{A520_327_B_C} (top right
panels) and their flux at 325 and 1400\,MHz along with their
integrated spectral indexes are given in Table\,\ref{fluxes of radio
  galaxies}.  Just three of them have been found to have an optical
counterpart, i.e. A, D and E that have been identified as cluster members (see Cooray et al. 1998 and Girardi et al. 2008).

The sources D and E are narrow angle tail radio galaxies with size $\sim$300
and 500\,kpc, located in the East of the cluster and oriented in the
opposite direction respect to the cluster center. Their spectra are
quite similar: flatter near to the nucleus ($\alpha\sim0.7$) and
steeper moving toward the external part of the tail, up to values 1.5
and higher.

Source A is a point-like source. According to the flux measured at
1400\,MHz and the upper limit inferred from the 325\,MHz image, we can
conclude that it is characterized by an inverted spectrum.

Sources B and F do not show an optical counterpart. The radio source F
is in the South-East of the cluster and its signal is under the noise
level at 325\,MHz, allowing only an upper limit on its
spectral index. Source B has a double lobed structure, with size $\sim$300\,kpc  assuming the redshift of A520,
located in the West of the cluster and has a spectral index ranging
across the source between 0.7 and 1.4.

\begin{figure}[ht]
  \centering
  \includegraphics[width=7cm, angle=0]{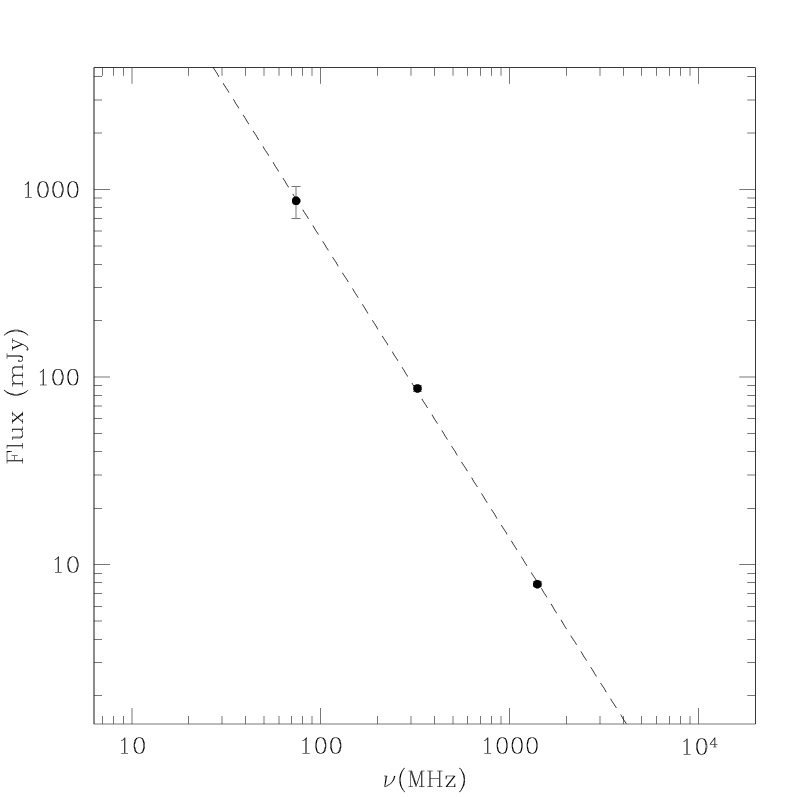}
  \caption{Spectral index of the source C (see e.g. Fig.\,\ref{A520_1.4_all}, top right panel) between 74 and 1400\,MHz. }
  \label{Spix_C}
\end{figure}

Source C is a powerful source in the North, only marginally
resolved at a resolution of $\sim$15$^{\prime\prime}$. From
Fig.\,\ref{A520_1.4_all} top right panel, two optical candidates slightly
displaced from the peak of the radio source emission can be
identified. This source is characterized by the steepest spectrum in
the field.  Between 325 and 1400\,MHz its spectral index is
$\alpha$=1.59$\pm$0.03. We investigated its emission also at lower
frequencies by using \emph{VLA Low-Frequency Sky Survey Redux} (VLSSr,
Lane et al. 2012). 
From the VLSSr image at 74\,MHz we measure a flux
$S_{\rm 74\,MHz}=(870\pm190)$\,mJy. This flux is expressed according to the Scaife \& Heald (2012) scale and has been corrected for the clean bias.  In Fig.\,\ref{Spix_C} the
spectrum of the source between 74 and 1400\,MHz is shown. From the fit of
the spectrum in this range of frequencies we infer an overall averaged spectral index
$\alpha$=1.60$\pm$0.03 in good agreement with the estimate obtained by
using only higher frequency data.  New JVLA observations at higher resolution
(Vacca et al. in preparation) reveal that the source is
actually a double-lobed radio source, while its emission is under the
noise level at 5000\,MHz.  The steep spectrum, the possible displacement
between the radio and optical emission and the double lobed structure
of this source suggest that it could be classified as a dying source (e.g. Parma et al. 2007, Murgia et al. 2011).
Alternatively, this source could be the radio counterpart of an high-redshift typical cluster-member elliptical-galaxy not detected in optical surveys because of sensitivity reasons.
\begin{table*}
\caption{Fluxes of radio galaxies}
\label{fluxes of radio galaxies}      
\centering          
\begin{tabular}{c c c c}
\hline\hline       
Label & $S_{\rm 325\,MHz}$ & $S_{\rm 1400\,MHz}$ & $\alpha_{\rm 325\,MHz}^{\rm 1400\,MHz}$ \\
&mJy&mJy\\
\hline
A&$\lesssim$3.3&6.4$\pm$0.2&  $\lesssim$-0.5   \\
B&19$\pm$1&5.2$\pm$0.2& 0.89$\pm$0.06\\
C&84 $\pm$3&8.2$\pm$0.3& 1.59$\pm$0.03\\
D&88$\pm$3&23.2$\pm$0.7&0.91$\pm$0.03\\
E&90$\pm$3&21.7$\pm$0.7& 0.97$\pm$0.03\\
F&$\lesssim$3.0&1.3 $\pm$  0.1&$\lesssim$0.6\\
\hline    
 \multicolumn{3}{l}{Col.\,1: Radio galaxy label; Col.\,2: Flux at 325\,MHz;}\\
\multicolumn{3}{l}{Col.\,3: Flux at 1400\,MHz; Col.\,4: Spectral}\\ 
\multicolumn{3}{l}{index between 325 and 1400\,MHz.} \\
\end{tabular}
\end{table*}

\section{Radio halo brightness profile}

According to Murgia et al. (2009), the azimuthally averaged brightness profile of radio halos can be modeled with an exponential law: 
\begin{equation}
I(r)=I_0e^{-\frac{r}{r_{\rm e}}}
\end{equation}
where $I_0$ and $r_{\rm e}$ are the central radio surface brightness and the $e$-folding radius. From the central brightness and the $e$-folding radius, the radio emissivity of the diffuse emission can be inferred. This model does not account for the local deviations from the
circular symmetry of the diffuse emission but gives a good description 
of averaged properties, involving a minimum set of free parameters and allowing a comparison with other radio halos. Murgia et al. note that it is a simplistic description and consider also more realistic pictures, as for example the injection model. 
According to this injection model, the radio halo brightness profile would be:
\begin{equation}
I(r_{\perp})=I_0\left(1+\frac{r^2_{\perp}}{r^2_{\rm c}}\right)^{-6\beta\eta+0.5}
\label{injectionbrightness}
\end{equation}
where $I(r_{\perp})$ is the brightness value at the $r_{\perp}$ projected distance from the cluster center and $I_0$ is the central radio halo brightness.
The diffuse emission is supposed to be generated by an isotropic population of relativistic electrons with a power law energy spectrum and in a steady state of continuous injection. A radio halo spectral index $\alpha=1$ is considered, corresponding to a perfect equipartition condition between particles and field (see Brunetti et al. 1997, Beck \& Krause 2005). The magnetic field has been assumed to be completely tangled and an average over all the possible directions between the magnetic field and the line-of-sight has been performed. The magnetic field strength is assumed radially decreasing as a function of the thermal gas density, taken to follow a $\beta$-model (Cavaliere \& Fusco-Femiano 1976).

By using the procedure implemented in the software FARADAY (Murgia et al. 2004), we tried to fit the azimuthally averaged radial profile of the radio halo in A520 both considering an exponential and an injection model. 
For the injection model, we adopted the gas density parameters derived by Govoni et al. (2001) from ROSAT X-ray data and re-scaled to our chosen cosmology ($n_{\rm e}(0)=3.8\times10^{-3}$cm$^{-3}$, $r_{\rm c}=127^{\prime\prime}=413$\,kpc and $\beta=0.87$).
In both cases the fits do not allow a good description of the data points.
Indeed, the radial profile keeps flat up to distances from the cluster center $\sim$400\,kpc, see Fig.\,\ref{A520_spix_radialprofile_beam}. This peculiarity has been already noticed by Govoni et al. (2001) who do not exclude that this emission could actually be a peripheral source seen in projection towards the cluster center.

\section{Discussion}
\label{Discussion}

\subsection{Radio halo spectral index image}
The spectral index image of the radio halo in A520 (Fig.\,\ref{Spix_all}) appears clumpy, with fluctuations both in the 39$^{\prime\prime}$ and in the 60$^{\prime\prime}$ resolution images.
The distribution of the spectral index has a dispersion slightly larger than the mean value of the spectral index error image. This suggests that the measurement process significantly contributes to the observed fluctuations even if a certain degree of intrinsic clumpiness seems to be present. 
According to primary models of radio halo formation (e.g. Schlickeiser et al. 1987; Petrosian 2001; Brunetti \& Lazarian 2007, 2011), magneto-hydro-dynamical turbulence due to cluster mergers is supposed to stochastically (re-)accelerate preexisting electrons through Fermi-II processes causing diffuse large scale synchrotron radio emission in massive galaxy clusters. In this scenario,  a complex cluster distribution of the spectral index is expected. On the contrary, secondary models of radio halo formation (e.g. Dennison 1980, Pfrommer et al. 2008, Ensslin et al. 2011) propose that the diffuse large scale synchrotron radio emission is due to a continuous generation of relativistic electrons through the collision between relativistic protons (accelerated during the cluster history) and thermal protons in the ICM. In this scenario, a uniform distribution of the radio halo spectral index is expected.

The intrinsic complexity observed in the radio halo spectral index image of A520 appears in agreement with the predictions from primary models, although secondary models can not be excluded. 
According to primary models a radial steepening and a complex spatial distribution of the spectral index is expected because of a different (re-)acceleration efficiency in different cluster regions and/or variations in the local magnetic field strength. Consequently, the presence of radial steepening in radio halo spectral index distribution would suggest that the outer regions of the radio halo are characterized by inefficient acceleration processes and/or the presence of a radial gradient in the magnetic field strength. 
Therefore, if interpreted in the context of a primary model scenario, the flatness in the spectral index of the radio halo in A520 could indicate a fine-tuning between the magnetic field strength and the power supply up to large ($\sim1$\,Mpc) distances from the cluster center, in the radio halo periphery.

\subsection{Shock wave}
Feretti et al. (2004) found that the spectral index in A665 steepens at increasing distances from the cluster center, along the direction of propagation of the shock. The authors interpret this behavior as due to the fact that shocks associated with major mergers are not strong enough to accelerate particles, as predicted by Gabici \& Blasi (2003). The shocks in A665 and in A520 are characterized by the same Mach number ($M\sim 2$), therefore we would expect in A520 a similar behavior as observed in A665. 
On the contrary, in the hypothesis that major merger shocks are efficient in accelerating radiating electrons and since no turbulence or other kind of acceleration mechanisms are supposed to be present in between the shock and the North-East clump, Markevitch et al. (2005) expect different spectral index for these two regions and a steepening of the spectral index along the shock in A520, starting from a value of $\alpha=1.2$ at the bow shock location.

By considering a sector along the direction of propagation of the shock and the cool tail in the NE, we measured a spectral index of $\sim$1.2 that keeps flat.
Even if we measure a spectral index $\alpha=$1.2 at the bow shock location in agreement with the predictions from Markevitch et al. (2005), our analysis does not reveal the steepening of the spectral index in the direction of the propagation of the shock that they expect. The 1$\sigma$ errors on spectral index rule out spectral steepening from SW to NE at a confidence level of $\sim$80\%. Moreover, the significant spectral difference between the radio emission at the shock location and that in the NE that Markevitch et al. (2005) suppose seems not to be present. 

The behavior of the spectral index distribution in A520 differs from that observed in A665 and indicates that some efficient powering mechanism is at work even at  large ($\sim$1\,Mpc) distances from the cluster center. 
The spectral slope observed at the shock location in A520 does not exclude shock-accelerated electrons, while the flatness of the spectrum along the direction of propagation of the shock and the absence of significant spectral differences between the SW and the NE of the diffuse emission could indicate the presence of turbulence in the intermediate region. An alternative possibility is that turbulence is powering the emitting particles up to distances of 1\,Mpc from the cluster center. 

Deep radio observations at higher resolution both at 1400 and at 325\,MHz would allow to measure the spectral index distribution directly behind the location of the shock with an high degree of accuracy, giving precious information to distinguish between the two scenarios.  Moreover, sensitive X-ray observations would allow to probe the possible presence of turbulence in the intermediate region and at $\sim$\,Mpc distance from the cluster center (see e.g. Schuecker et al. 2004 for the Coma cluster).

\subsection{Radio halo spectral index versus thermal gas temperature}

To infer some information about a possible correlation between the radio and X-ray emission in the cluster, we compared the spectral index distribution  with the thermal gas temperature image. 
 
A520 follows the trend of the integrated spectral index versus temperature observed for other clusters. 
A strong point-to-point correlation between the radio-halo spectral index and the thermal-gas temperature is not present in A520, see Fig.\,\ref{Temperature}.
A similar case has been observed in MACSJ\,00717$+$3745 (Bonafede et al. 2009a). In that case the authors conclude that it is due to projection effects. 
A correlation between the distribution of the radio halo spectral index and the thermal gas temperature has been observed in the galaxy cluster A2744 (Orr\`u et al. 2007). In the case of A520, the absence of such a correlation could be due to the peculiarity of the diffuse emission and to the fact that it could actually be a source at the cluster periphery seen in projection, as Govoni et al. (2001) noted, see the discussion in \S\,\ref{radio halo profile}.

\subsection{Radio halo nature}
\label{radio halo profile}
The diffuse synchrotron emission in A520 has been originally classified as a cluster radio halo because of its location (Giovannini et al. 1999).
The high resolution images of the radio halo at 1400\,MHz (Fig.\,\ref{A520_1.4_all}, top right and bottom panels) show that the brighter structures of the diffuse emission are located in the North-East of the cluster, at the same spatial location of the cool tail, and in the South-West of the cluster, aligned with the shock. In between, a depression in the radio surface brightness can be observed. A clear link is present between the mass distribution (Clowe et al. 2012) and the diffuse radio emission. Indeed, we note that the non-thermal emission permeates regions of high mass.
The NE clump is aligned with the elongated mass distribution (NE to SW): structures 2+3 in Fig.\,2 by Clowe et al. (2012).
The diffuse emission at the shock location in the SW is observed in correspondence of the 4-7 structures in the same figure.

In contrast with most of the radio halos observed to date 
the azimuthally averaged brightness profile can not be modeled either 
with an exponential law or with a more realistic model based on the
continuous injection of relativistic electrons (Murgia et al. 2009). 
Indeed, the brightness remains flat up to distances $\sim$400\,kpc from the cluster center. 
This flatness has been already noted by Govoni et al. (2001) who 
do not exclude that this is a source at the cluster periphery, 
apparently located at the cluster center because of projection effects.
Another possible explanation could be that we are looking at a relic in the SW and a young still forming radio halo in the NE. In the past the cluster could have been crossed by several shocks, one of them is now powering the diffuse non thermal emission in the SW. Along the tail of the shock, turbulence due to the merger is allowing the formation of a radio halo. This would explain the absence of central peak in the radio emission. A similar emission has been observed in the cluster 1RXS\,J0603.3+4214 (van Weeren et al. 2012).

Typically, elongated relics are characterized by a filamentary structure that has not been observed in this case with the available data. Sensitive high resolution observations are necessary to rule out the presence of filamentary substructures. The spectral index of the emission in the SW edge does not show any obvious distribution. Elongated relics show a clear spectral steepening toward the cluster center, while a specific trend has not been detected in roundish relics (see Feretti et al. 2012). 
At present we do not have information about polarization properties of the source. The degree of polarization of the emission could be very useful in testing the relic scenario. A work where polarimetric properties will be analysed and discussed is currently in progress (Vacca et al. in preparation).

\section{Conclusions}
\label{Conclusions}
In this paper we presented the spectral index image of the radio halo at the center of the cluster A520, obtained by comparing VLA observations at 1400 and 325\,MHz.
We found an integrated spectral slope $\alpha_{\rm 325}^{\rm 1400}=1.12\pm0.05$.

The clumpy morphology in the radio halo spectral index image supports the primary models of radio halo formation. 
In this framework, the flatness of the spectrum could suggest an ongoing merger phenomenon that uniformly and continuously (re-) accelerates relativistic electrons \emph{in situ} up to the distance where the shock wave has been detected.

The integrated spectral index and the global thermal gas temperature are in good agreement with the trend observed in other galaxy clusters, although we do not observe a clear point-to-point correlation between the spectral index and the thermal gas temperature. 
The lack of this correlation can be explained in the context of the peculiar nature of the radio emission at the center of the cluster. This emission has been originally classified as a radio halo but the brightness profile of this source strongly differs form those observed in other radio halos.
New deep polarimetric observations over a wide frequency range are necessary to better understand the nature of the diffuse emission in A520 and to investigate the connection between the thermal and non-thermal properties of this galaxy cluster.

\begin{acknowledgements}
We thank the referee for helpful comments and suggestions that improved the paper. The research was partially supported by PRININAF2009.  The National
Radio Astronomy Observatory (NRAO) is a facility of the National
Science Foundation, operated under cooperative agreement by Associated
Universities, Inc. Basic research in radio astronomy at the Naval Research Laboratory is supported by 6.1 Base funding. 
\end{acknowledgements}

\end{document}